\documentclass{iaus}
\usepackage{graphicx}

\title[~~Extragalactic Background Light] %% give here short title %%
{Spectrophotometric measurement of \\ the Extragalacic Background Light}

\author[K. Mattila, K. Lehtinen, P. V\"ais\"anen, G. von Appen-Schnur \& Ch. Leinert]   %% give here short author list %%
{Kalevi Mattila$^1$, 
Kimmo Lehtinen$^1$, 
Petri V\"ais\"anen$^2$,\\ Gerhard von Appen-Schnur$^3$, 
and Christoph Leinert$^4$}

\affiliation{$^1$ University of Helsinki, FI-00014 Helsinki, Finland, 
email: {\tt mattila@cc.helsinki.fi} \\
[\affilskip]$^2$ South African Astronomical Observatory and SALT, Cape Town, South Africa \\
[\affilskip]$^3$ Astronomisches Instititut, Ruhr-Universit\"at-Bochum, D-44801 Bochum, Germany \\
[\affilskip]$^4$ Max-Planck-Institut f\"ur Astronomie, D-69117 Heidelberg}

\pubyear{2011} %% 
\volume{284}  %% insert here IAU Symposium No.
\pagerange{1--12}
\jname{The Spectral Energy Distribution of Galaxies}
\editors{R.J. Tuffs \&  C.C.Popescu, eds.}
\begin{document}

\maketitle

\begin{abstract}
The Extragalactic Background Light (EBL) at UV, optical and NIR
wavelengths consists of the integrated light of all  unresolved
galaxies along the line of sight plus any contributions by intergalactic
matter including hypothetical decaying relic particles.
The measurement of the EBL has turned out to be a tedious problem.
This is because of the foreground components of the night sky brightness,
much larger than the EBL itself: the Zodiacal Light (ZL), Integrated
Starlight (ISL), Diffuse Galactic Light (DGL) and, for ground-based
observations, the Airglow (AGL) and the tropospheric scattered light.
We have been developing a method for the EBL measurement which
utilises the screening effect of a dark nebula on the EBL. A differential
measurement in the direction of a high-latitude dark nebula and
its surrounding area provides a signal that is due to two components
only, i.e. the EBL and the diffusely scattered ISL from the cloud.
We present a progress report of this method where we are now utilising
intermediate resolution spectroscopy with ESO's
VLT telescope. We detect and remove the scattered ISL component  by
using its characteristic Fraunhofer line spectral signature. In contrast
to the ISL, in the EBL spectrum all spectral lines are washed out. 
We present a high quality spectrum
representing the difference between an  opaque position within 
our target cloud and several clear OFF positions around the cloud. 
We derive a preliminary EBL value at 400 nm and an upper limit to the EBL
at 520 nm. These values are in the same range as the EBL lower limits
derived from galaxy counts. 

\noindent{\bf Unit:} We will use in this paper the abbreviation 
1 cgs = $10^{-9}$erg\,s$^{-1}$cm$^{-2}$sr$^{-1}$\AA$^{-1}$\,


%% add here a maximum of 10 keywords, to be taken form the file <Keywords.txt>
\end{abstract}

\firstsection % if your document starts with a section,
              % remove some space above using this command.
\section{Introduction}

The importance of the Extragalactic Background Light 
(EBL) for cosmology has long been recognized 
(e.g. \cite[Partridge \& Peebles 1967]{Partrige67}). 
The EBL at UV, optical and near infrared wavelengths consists
of the integrated light of all unresolved  galaxies along the line of sight
plus any contributions by intergalactic gas and dust and by hypothetical
decaying relic particles. A large fraction of the energy released in 
the Universe since the recombination epoch is expected to be contained
in the EBL. An important aspect is the balance between the UV-NIR 
and the far infrared EBL: what is lost by dust obscuration
in the UV-NIR will re-appear in the FIR. 
Some central,
but still largely open astrophysical problems to be addressed through EBL
measurements include the formation and early evolution of galaxies and
the star formation history of the Universe. 
Because of the foreground components,
much larger than the EBL itself, the measurement of the EBL has
turned out to be a tedious problem. As
a consequence we lack a generally accepted measured value of the 
optical EBL. For a review see Leinert et al. (1998). 

Recently, \cite[Bernstein et al. (2002)]{Bernstein02_I} 
have announced ``the first detection'' 
of the EBL at 300, 550, and 800 nm. They used a combination of 
space borne (HST) and ground based (Las Campanas) measurements in their 
method. However, \cite[Mattila (2003)]{Mattila03} has argued that they 
neglected important effects of the 
atmospheric scattered light and had a problem with the inter-calibration 
of the two telescopes used. Therefore, the claim for a detection of 
the EBL appeared premature. After reanalysis of their systematic errors
\cite[Bernstein et. al. (2005)]{Bernstein05} and \cite[Bernstein 
(2007)]{Bernstein07} came to the same conclusions which they formulated 
as follows:
   ``  ... the complexity of the corrections required to do absolute surface 
    (spectro)photometry from the ground make it impossible to achieve 
    1\% accuracy in the calibration of the ZL.'' and
   ``...the only promising strategy ... is to perform all measurements 
   with the same instrument, so that the majority of corrections and 
   foreground subtractions can be done in a relative sense, 
   before the absolute calibration is applied.''

In this situation it is highly desirable 
to obtain a measurement of the EBL with an independent method which
utilises one and the same instrument for all sky components and is 
virtually free of the large foreground components, the Airglow (AGL), 
Zodiacal Light (ZL) and tropospheric scattered light.

\section{The dark cloud shadow method.}

We have been developing over several years a method for the measurement
of the EBL which utilises the screening effect of a dark nebula on the
background light (see Mattila 1990 for a review and previous results).
A differential measurement of the night-sky brightness in the direction
of a high galactic latitude dark nebula and its surrounding area,
which is (almost) free of obscuring dust, provides a signal that is
due to two components only: (1) the EBL and
(2) the diffusely scattered starlight from interstellar dust in the 
cloud (and to smaller extent also in its surroundings). All the
large foreground components, i.e. the ZL, the AGL, and the tropospheric 
scattered light, are completely eliminated (see Fig. 1a). 
The direct starlight down to $\sim$21-23 mag
can be eliminated by selecting the measuring areas with the help
of deep images.
At high galactic latitudes ($|b|>$30 deg), the star density is 
sufficiently low to allow blank areas of sufficient size to be easily
found. Models of the Galaxy  show that the contribution from unresolved stars
beyond this limiting magnitude is of minor or no importance (Mattila 1976).
If the scattered light from the interstellar dust were zero (i.e.
if the grain albedo $a = 0$), then the difference
in surface brightness between a transparent comparison area and the 
dark nebula would be due to the EBL only, and an opaque nebula
would be darker by the amount of the EBL intensity (dashed line
in Fig. 1c). The scattered light is not zero, however.
A dark nebula in the interstellar space is always exposed to the 
radiation field of the integrated Galactic starlight, which gives
rise to a diffuse scattered light (shaded area in Fig. 1c). Because
the intensity of this scattered starlight in the dark nebula is
equal to or larger than the EBL, its separation is the main task
in our method.  The key issue in the dark cloud method is to get
a reliable estimate for the scattered light.
This can be achieved by means of spectroscopy.
With intermediate-resolution spectroscopy ($R \approx 800$) 
it is possible to determine the strengths of the stronger Fraunhofer lines in
the spectrum of the excess surface brightness of the dark cloud.
Spectroscopic measurement of a faint surface brightness
signal, only 1 -- 10\% above the dark sky level has been possible with the 
VLT/FORS long slit spectroscopy.
\smallskip

{\em Target cloud and positions.} The high galactic latitude dark nebula Lynds~1642

\begin{minipage}{6cm}
($l =  210.9\deg, b = -36.7\deg$) 
has been chosen as our primary target. 
It has a high  obscuration ($A_V > 15$ mag)
in the centre and areas of good transparency ($A_V \approx 0.15$ mag) in its 
immediate surroundings within $\sim 1-2 \deg$. Its declination of $-14.5\deg$ 
allows observations at low airmasses from Paranal. Our measuring positions
($2' \phi$) for intermediate band photometry (Mattila and Schnur 1990) are shown
in Fig.~2 as circles and the positions where photometry and VLT/FORS  
long slit spectra (2''x6.8') were taken, as squares. 
Differential measurements
with repeated ON/OFF switching cycle of $\sim$ 0.5 h were used to eliminate 
airglow variations. The spectral range was $360 -600$ nm and the resolution  
$\Delta \lambda = 1.1$ nm. Stamps of 10'x10' size 
centered on the VLT/FORS positions, adopted from DSS blue plates, are shown in the margins:  
the dark central position \#8 ($A_V > 15$ mag)  in upper left, followed clockwise by the two
intermediate opacity positions \#9 and 42, and then the transparent OFF positions. Two observed spectra ``dark nebula minus surrounding sky'' are
shown in Fig.~3:  One is for the opaque central position \#8 with $A_V > 15$ mag,
the other for the average of two intermediate opacity
\end{minipage}
\hfill
\begin{minipage}{5.5cm}
\setlength{\unitlength}{1mm}
\begin{picture}(0,0)
\put(0,0){\begin{picture}(0,0) \includegraphics{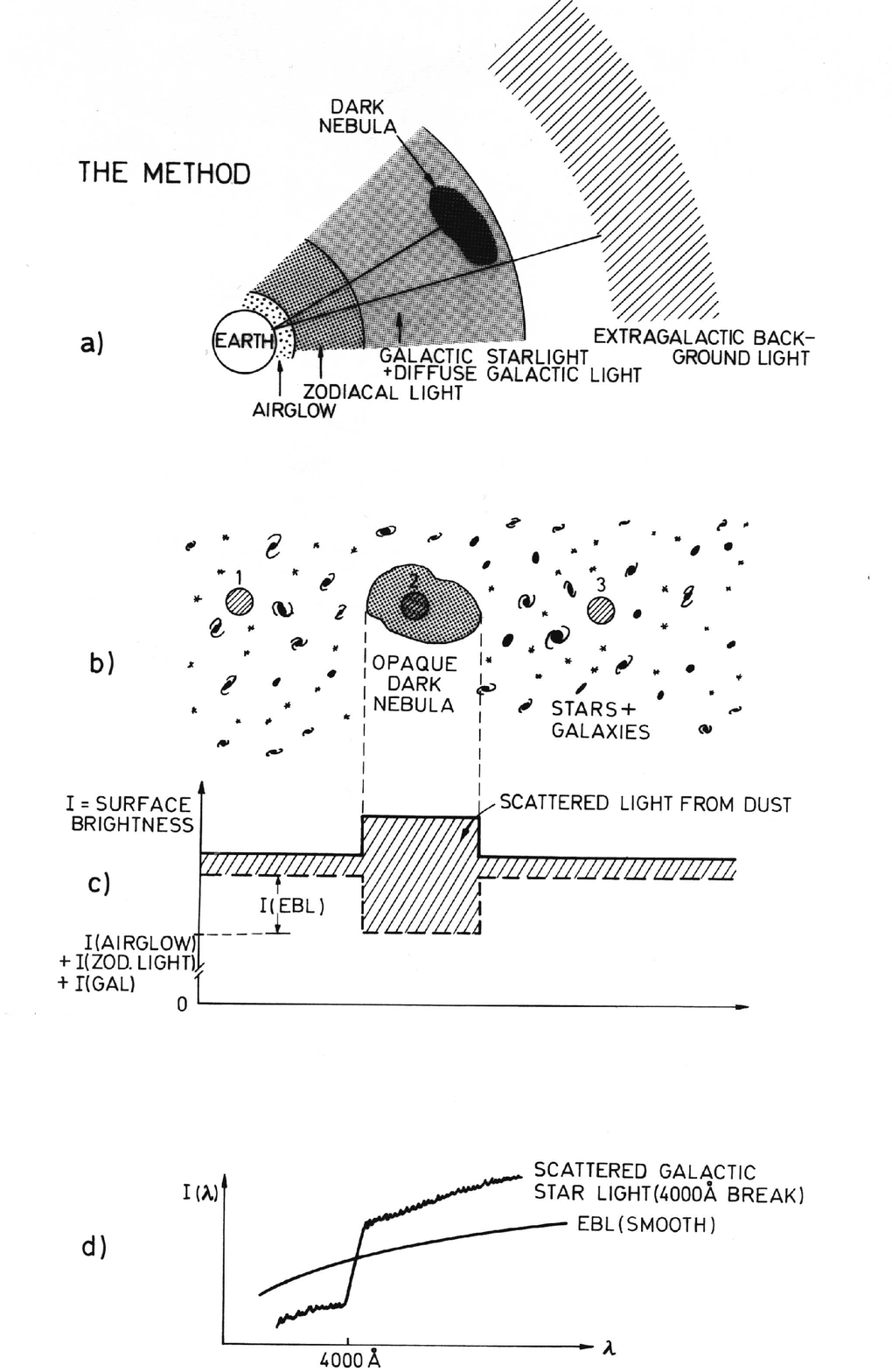} \end{picture}}
\put(30,-60){\makebox(0,0) {\small {\bf Figure 1.} EBL measurement with}}
 \put(30,-63){\makebox(0,0) {\small the dark cloud shadow method}} 
\end{picture}
\end{minipage}

\vspace{0.1cm}

\noindent   positions, \#9 and \#42
with  $A_V \approx$ 1~mag. The crosses show the results of previous intermediate
band photometry by Mattila and Schnur (Mattila 1990) with error bars and filter 
band widths indicated. The spectra have been re-scaled by a multiplicative factor
to adjust them to the more reliably calibrated photometry. 
The intermediate opacity positions               
are used to validate the model of Galactic starlight spectrum, 
needed to accurately separate the Galactic contribution from the observed 
spectrum  ``dark position minus clear sky''.
In combination with the dark central position they
 enable us to disentangle the effects of extinction and reddening by dust
from the spectral shape of the Galactic starlight.
 
\smallskip
{\em Component separation method.}
 Our separation method utilises the difference 
in the spectra of the EBL  and the integrated Galactic starlight (see Fig 1d).
While the scattered Galactic starlight spectrum has
the characteristic stellar Fraunhofer lines and the  
discontinuity at 400 nm the EBL spectrum is a smooth one without 
these features. This can be understood because the radiation from galaxies
and other luminous matter over a vast redshift range contributes
to the EBL, thus washing out any spectral lines or discontinuities.
The spectrum of the integrated Galactic starlight can
be synthesised by using the known spectra of representative stars
of different spectral classes, as well as the

\newpage

\begin{minipage}{13.5cm}
\setlength{\unitlength}{1mm}
\begin{picture}(0,0)
\put(0,0){\begin{picture}(0,0) \includegraphics{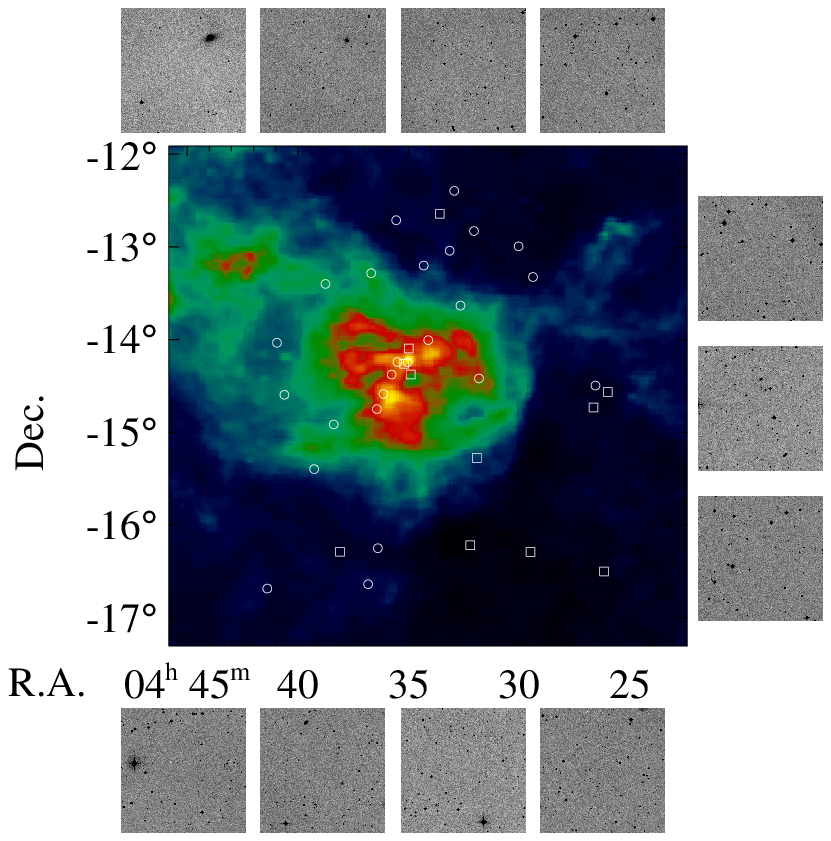} \end{picture}}
\put(50,-120){\makebox(0,0) {\small {\bf Figure 2.} The observed positions in the L~1642 cloud area.}}
\put(50,-123){\makebox(0,0) {\small superimposed on an IRAS 100 $\mu$m map. For details see text.}} 
\end{picture}
\end{minipage}

\begin{minipage}{10.5cm}
\setlength{\unitlength}{1mm}
\begin{picture}(0,0)
\put(0,0){\begin{picture}(0,0) \includegraphics{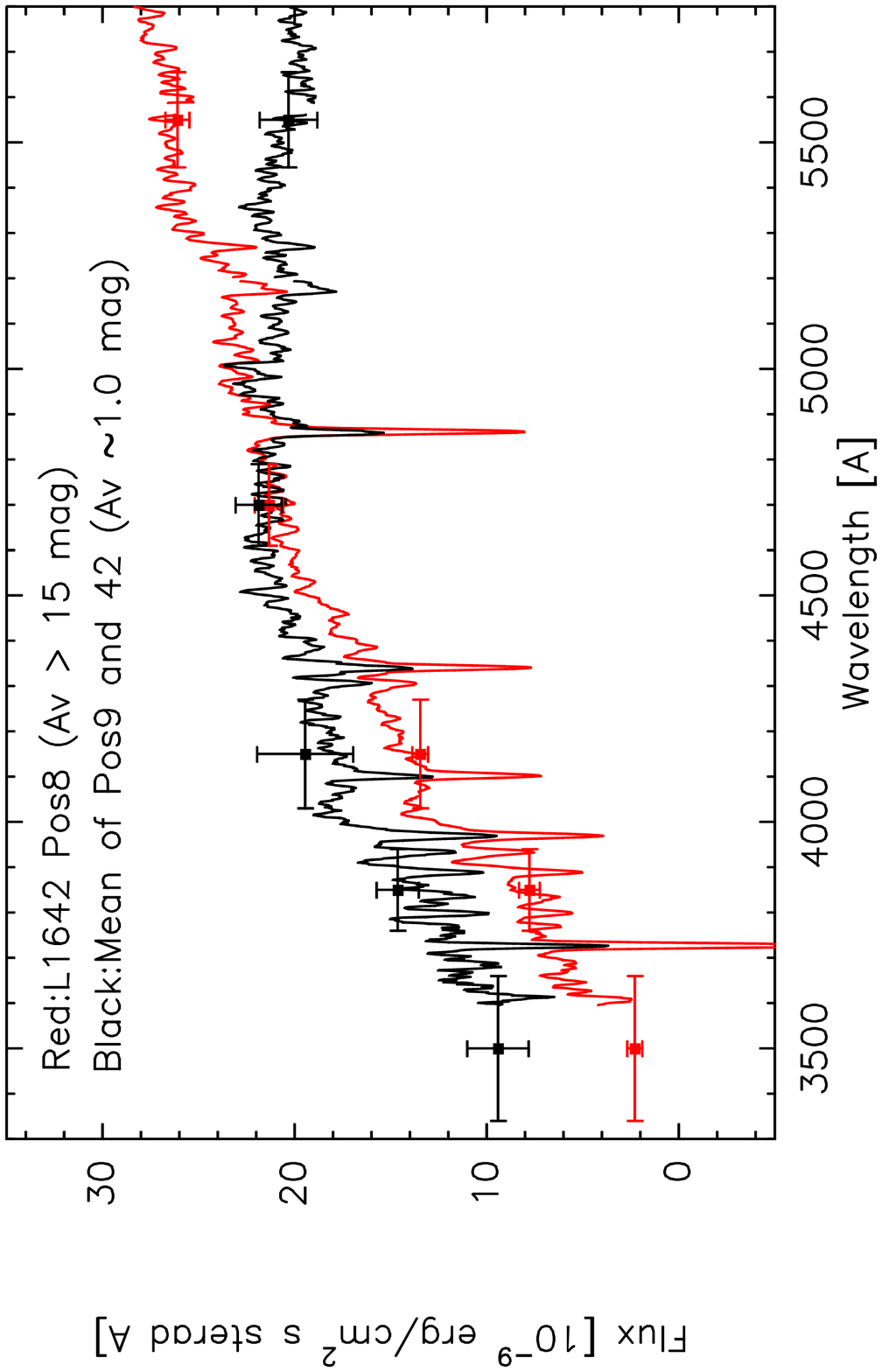} \end{picture}}
\put(60,-196){\makebox(0,0) {\small {\bf Figure 3.}Observed spectrum ``dark nebula minus surrounding sky'' for the central}}
 \put(60,-199){\makebox(0,0) {\small  opaque position \#8 and the mean of two translucent positions, \#9 and \#42.}}
 \put(60,-199){\makebox(0,0) {\small }} 
\end{picture}
\end{minipage}

\newpage

\noindent  distribution of stars and dust in the Solar neighbourhood. Such synthetic spectra have been calculated using the model of Mattila 
(1980) and the recent STELIB library of 
stellar spectra (Le Borgne et al. 2003).
The resulting ISL spectrum, mean over sky, is shown in Fig.~4 (the uppermost
blue line). This spectrum, 
smoothed to our FORS resolution of 1.1 nm, scaled and reddened to correspond
to the dark cloud scattered light spectrum, shows several strong 
Fraunhofer lines and the discontinuity at 400 nm. 
For the separation of the scattered light we cannot, however, rely on
the integrated starlight {\em model} alone. We have to test and validate
this model by observations at intermediate opacity positions of the 
dark cloud where most of the EBL is transmitted through the cloud and contributes
only little to the surface brightness difference ``dark cloud minus surroundings''.

\section{Separation of the scattered light}

 The observed surface brightness difference, $\Delta I$,
``dark nebula minus surrounding sky'' consists of the following main components
(followed by {\bf +} or {\bf -} depending on whether it makes a positive or negative 
contribution to $\Delta I$): {\bf (1)} Starlight scattered by dust in the nebula {\bf (+)};
{\bf (2)} Starlight scattered by dust in the comparison fields of the surrounding sky  {\bf (-)};
{\bf (3)} diffuse line emission by ionised gas in the comparison fields 
beyond the distance of the dark cloud  {\bf (-)};
{\bf (4)} the Extragalactic Background Light (present in the comparison fields) {\bf (-)}.

In order to derive an estimate for the EBL 
it is necessary to separate the contribution of scattered light (components 1 and 2). 
When applying the dark cloud method to intermediate band photometry, we 
utilised the 400 nm discontinuity which is strong and well defined in the
integrated starlight spectrum and is present also in the scattered light 
spectra as shown in Fig.~3. However, its value as determined using intermediate 
bands is strongly influenced by the wavelength-dependent extinction and multiple 
scattering in the cloud which need to be modeled.

We use our model ISL spectra and the observed VLT/FORS spectrum for the opaque 
central position \#8 to separate the scattered light component. 
The scattered light spectrum is assumed to be a copy 
of the ISL spectrum but to be reddened by a factor linearly
proportional to the wavelength. We show in Fig.~4 an overall model fit for 360 - 550 nm
assuming that $I(EBL) = 2$ cgs at OFF and = 0 at the \#8 ON position. 
The ``ON-position'' model spectrum (blue line) is 
the mean ISL spectrum over the whole sky, suitably scaled (by factor 0.18 at 400 nm) and
linearly reddened (by 0.075 per 100 nm). The ``OFF-position'' model spectrum (green line)
is the ISL spectrum for $|b| = 37$ deg, scaled according to the waveleghth dependent
extinction (determined from 2MASS JHK stellar and ISOPHOT 200 $\mu$m surface photometry).
The ON minus OFF model spectrum is shown as red line. The observed ON minus OFF
spectrum for Pos \#8 (as in Fig.~3) is shown as black line. As can be seen the fit is 
almost perfect except for the Balmer lines. These lines are contaminated by the excess
emission of diffuse ionised gas in the OFF positions. At this stage we do not attempt
to model this component (shown as magenta line in Fig.~4 on  arbirtrary scale)
and exclude the Balmer lines from our fitting procedure.

\section{EBL estimates from the dark cloud method}

Although a good fit of the observed spectrum by scattered ISL only can be 
achieved there are small differences in the depths of the
Fraunhofer lines (other than the Balmer Lines) indicating that $I(EBL) \ne 0$. 
We have made fits over suitably selected narrow wavelength intervals.  
The EBL is assumed to be constant over the wavelength 

\newpage

\begin{minipage}{11.5cm}
\setlength{\unitlength}{1mm}
\begin{picture}(0,0)
\put(0,0){\begin{picture}(0,0) \includegraphics{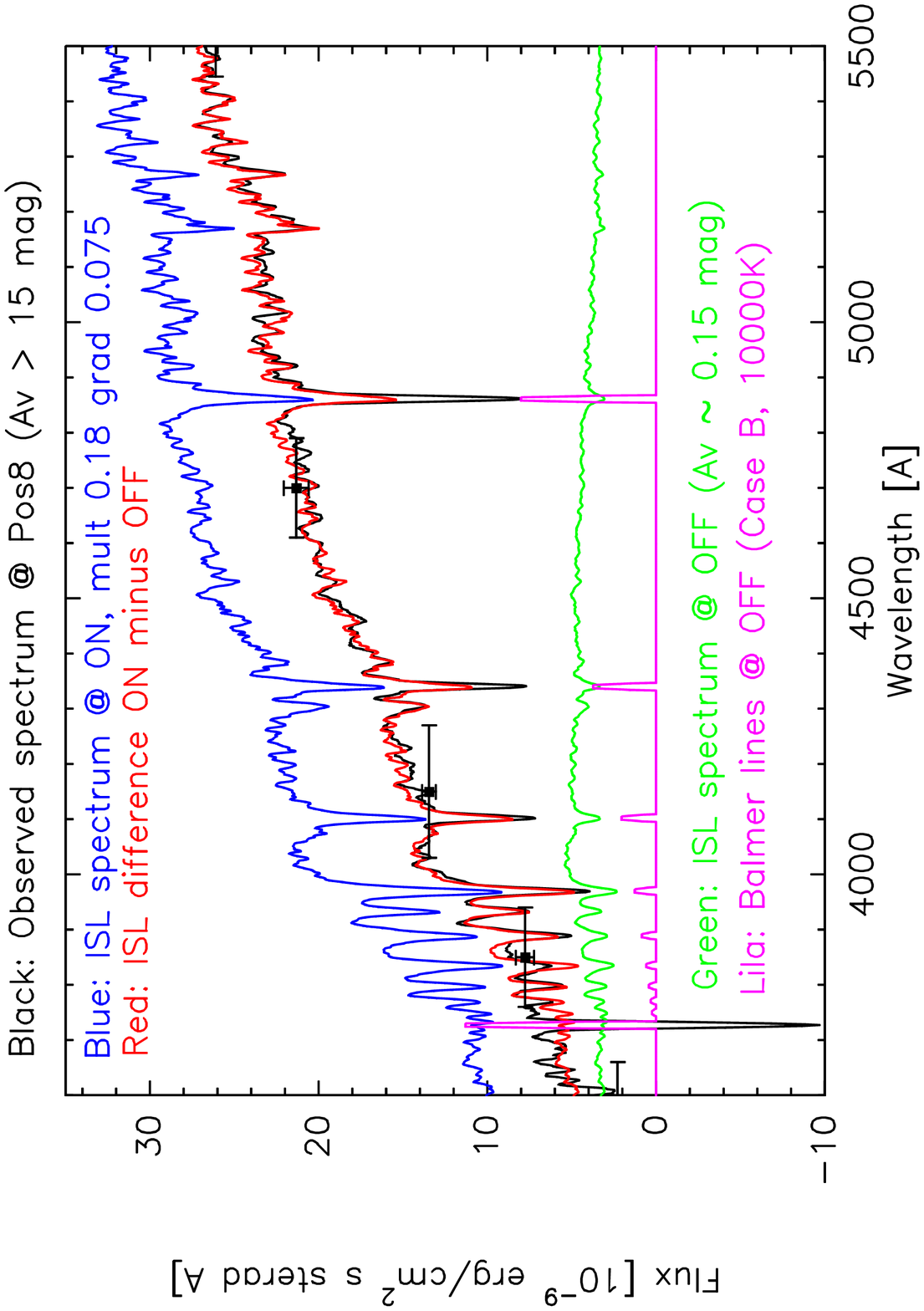} \end{picture}}
\put(70,-100){\makebox(0,0) {\small {\bf Figure 4.} Fitting the central 
opaque position \#8 spectrum with the ISL spectrum.}}
\put(70,-103){\makebox(0,0) {\small I(EBL) = 2 cgs has been assumed for OFF positions.
See text for details }}
%\put(50,-123){\makebox(0,0) {\small with the ISL spectrum}} 
\end{picture}
\end{minipage}

\begin{minipage}{8.5cm}
\setlength{\unitlength}{1mm}
\begin{picture}(0,0)
\put(0,0){\begin{picture}(0,0) \includegraphics{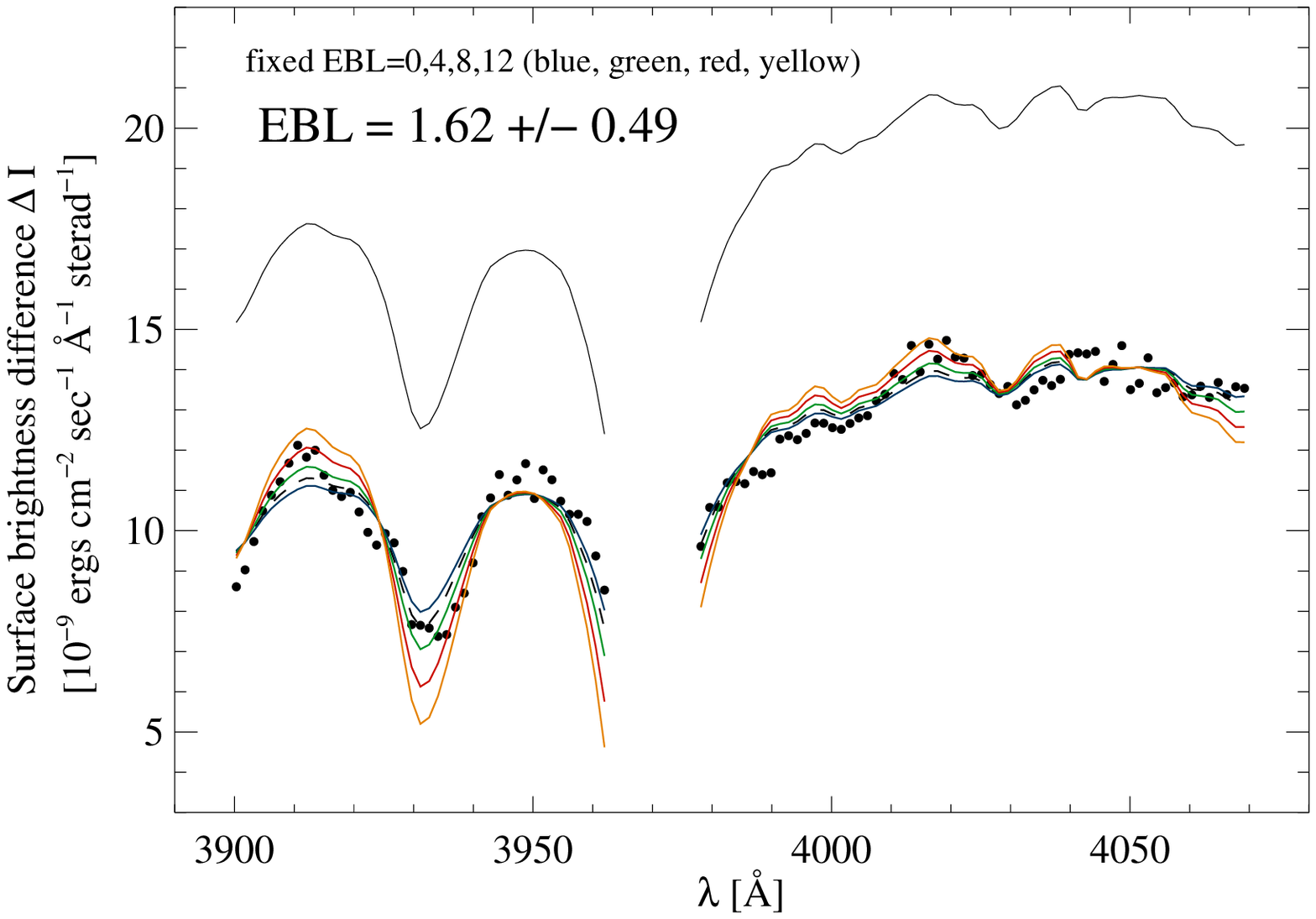} \end{picture}}
\put(50,-185){\makebox(0,0) {\small {\bf Figure 5.} Fitting of the 390 - 407 nm spectral range}}
\put(50,-188){\makebox(0,0) {\small of the opaque position 8 spectrum with the Integrated}}
\put(50,-191){\makebox(0,0) {\small Starlight spectrum and different assumed values}} 
\put(50,-194){\makebox(0,0) {\small of the EBL}} 
\end{picture}

\end{minipage}
\hfill
\begin{minipage}{4cm}
\setlength{\unitlength}{1mm}
\begin{picture}(0,0)
\put(0,0){\begin{picture}(0,0) \includegraphics{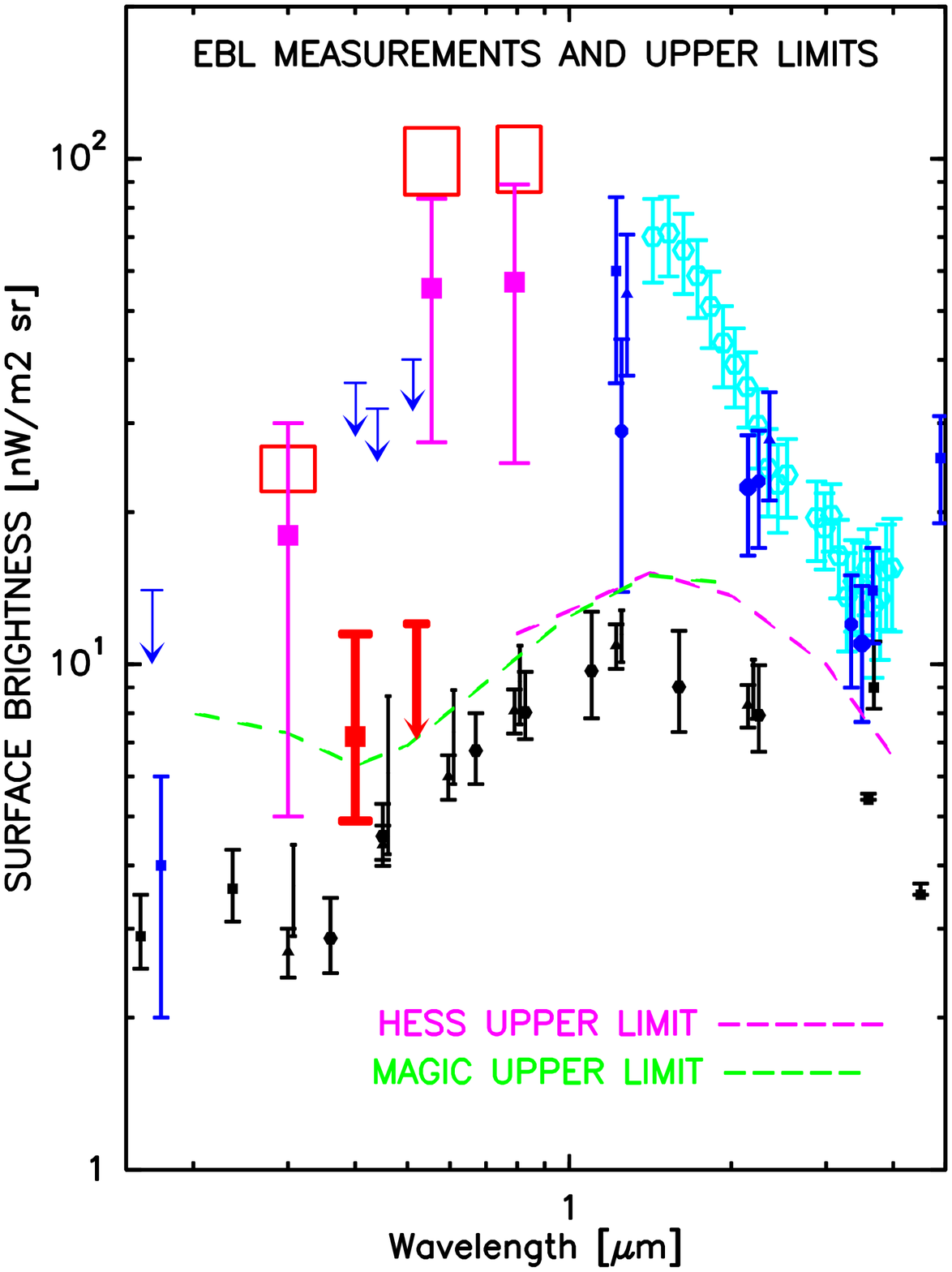} \end{picture}}
\put(30,-185){\makebox(0,0) {\small {\bf Figure 6.} Compilation of }}
\put(30,-188){\makebox(0,0) {\small current EBL measurements, }}
\put(30,-191){\makebox(0,0) {\small lower and upper limits. }}
\put(30,-194){\makebox(0,0) {\small See text for symbols}} 
\end{picture}
\end{minipage}

\newpage

\noindent slot in question.
The ISL scaling factor and wavelegth dependent gradient are free parameters. 
The fit to the range 390 - 407 nm is shown in Fig.~5. 
The Ca{\footnotesize II} H line at 396.9 nm was not included 
because it is contaminated by the H$\epsilon$ line of ionised gas. Our choice
for another suitable slot with strong Fraunhofer lines  was 510 - 535 nm.
By fitting the observed spectrum with an ISL spectrum and EBL contributions 
of different amounts we obtain the following EBL values for these two slots:\\
$I(EBL)$(400 nm) = $1.6 \pm 0.5$ cgs $(1\sigma)$ or $6.4 \pm 2.0$ nW m$^{-2}$ sterad$^{-1}$\\
$I(EBL)$(520 nm) $\le 1.6$ cgs $(2\sigma)$ or $\le 8.3$ nW m$^{-2}$ sterad$^{-1}$ 

The statistical error ($1\sigma$) is small enough for putting  to the EBL
significant limits comparable with the EBL minimum value of  
$\sim 0.8 - 1.0$ cgs for $\lambda = 360 - 810$~nm derived by summing up deep galaxy 
counts (Madau and Pozzetti 2000).

{\em Systematic errors.}
(1) Blocking by the dark cloud is not complete for incident isotropic diffuse radiation
like the EBL. 
For $A_V = 16$ mag the blocking factor is  0.86 (see Mattila 1976, Table A2).
(2) The systematic error caused by our ISL model is tested by 
observing the semi-transparent
positions with $A_V \approx 1$ mag for which the scattered starlight has the same 
spectrum but the EBL contribution is only $\sim 30\%$ of that for $A_V = 15$ mag. 
They give $I(EBL)\approx$ 0 (within 1$\sigma$ statistical errors) 
and may thus indicate a need to increase the EBL upper boundary error bar by 
$\sim 0.3 \times I(EBL)$ or $\sim$ 0.5 cgs.

With a 14\% blocking factor correction and increasing the upper
bound of the errors (systematic) by 0.5 cgs we end up with
the following preliminary EBL estimates:\\
$I(EBL)$(400 nm) = $1.8^{+1.0}_{-0.5}$ cgs $(1\sigma)$ or $7.2^{+4}_{-2}$ nW m$^{-2}$ sterad$^{-1}$\\
$I(EBL)$(520 nm) $\le 2.3$ cgs $(2\sigma)$ or $\le 12.0$ nW m$^{-2}$ sterad$^{-1}$

We show in Fig. 6 a compilation from UV to NIR (0.1 - 5 $\mu$m) of recent EBL 
measurements and upper limits (colour symbols) as well as lower limits 
from galaxy counts (black symbols). Upper limits from the TeV gamma-ray absorption 
method are shown as dashed lines. The resulting values from the present paper
are shown as red solid square with error bars at 400 nm and upper limit at 520 nm. 
The reanalysis by Mattila (2003) of the Bernstein et al. (2002) EBL values
are shown as large red rectangles, indicating upper limits. In agreement with 
these, the reanalysis by Bernstein (2007) resulted in values  that
were substantially increased from their original estimates.
They are shown as the solid magenta squares with error bars.
For references to the three upper limits at 400 - 520 nm see Leinert et al. (1998),
and for the other points see Mattila (2006) and Dominguez et al. (2011). For a
new EBL estimate (not shown in Fig.~6) based on reanalysis of Pioneer10/11 data 
see Matsuoka et. al. (2011) and these Proceedings.

{}

\end{document}